\documentstyle[aps,eqsecnum,12pt]{revtex}

\begin{document}

\begin{flushright}
OU-TAP-39 \\
\today
\end{flushright}


\begin{center}
{\Large \bf Formulation for nonaxisymmetric uniformly rotating 
equilibrium configurations in the second post-Newtonian 
approximation of general relativity}
\end{center}

\vspace{1cm}

\begin{center}
Hideki Asada\footnote{Electronic address: asada@vega.ess.sci.osaka-u.ac.jp} 
and Masaru Shibata\footnote{Electronic address: 
shibata@vega.ess.sci.osaka-u.ac.jp}
\end{center}

\begin{center}
Department of Earth and Space Science, Graduate School of Science \\
Osaka University,~Toyonaka, Osaka 560,~Japan
\end{center}

\vspace{1cm}

\begin{center}
{\bf ABSTRACT}
\end{center}

\begin{abstract}
We present a formalism to obtain equilibrium configurations of 
uniformly rotating fluid in the second post-Newtonian approximation 
of general relativity. In our formalism, we need to solve 29 Poisson 
equations, but their source terms decrease rapidly enough at the external 
region of the matter(i.e., at worst $O(r^{-4})$). 
Hence these Poisson equations can be solved accurately as the boundary 
value problem using standard numerical methods. 
This formalism will be useful to obtain nonaxisymmetric 
uniformly rotating equilibrium configurations such as 
synchronized binary neutron stars just before merging and the Jacobi 
ellipsoid. 
\end{abstract}

\vspace{5mm}

\begin{flushleft}
PACS number(s): 04.25.Nx
\end{flushleft}

\def\pa{\partial}
\def\bI{\hbox{$\,I\!\!\!\!-$}}
\def\two{\hbox{$_{(2)}$}}
\def\three{\hbox{$_{(3)}$}}
\def\four{\hbox{$_{(4)}$}}
\def\five{\hbox{$_{(5)}$}}
\def\six{\hbox{$_{(6)}$}}
\def\seven{\hbox{$_{(7)}$}}
\def\eight{\hbox{$_{(8)}$}}
\newcommand{\lsim}{\raisebox{0.3mm}{\em $\, <$} \hspace{-3.3mm}
\raisebox{-1.8mm}{\em $\sim \,$}}

\newpage
\baselineskip 8mm

\section{Introduction}

The last stage of coalescing binary neutron stars(BNS's) is one of the most 
promising sources for kilometer size interferometric 
gravitational wave detectors, 
LIGO\cite{ligo} and VIRGO\cite{virgo}. When the orbital separation of 
BNS's becomes $\sim 700$km as a result of the emission of gravitational 
waves, it is observed that the frequency of gravitational waves from them 
becomes $\sim 10$Hz. 
After then, the orbit of BNS's shrinks 
owing to the radiation reaction toward merging in a few minutes\cite{cult}. 
In such a phase, BNS's are the strongly self-gravitating bound systems, and 
gravitational waves from them will have various general relativistic(GR) 
imformations. 
In particular, in the last few milliseconds before merging, BNS's are 
in a very strong GR gravitational field because the orbital 
separation is less than ten times of the Schwarzschild radius of the system. 
Thus, if we could detect the signal of gravitational waves radiated in the 
last few milliseconds, we would be able to observe directly the phenomena 
in the GR gravitational field. 

To interpret the implication of the signal of gravitational waves, 
we need to understand the theoretical mechanism of merging in detail. 
The little knowledge we have about the very last phase of BNS's is 
as follows: When the orbital separation of BNS's 
is $\lsim 10 GM/c^2$, where $M$ is the total mass of BNS's, 
they move approximately in circular orbits because 
the timescale of the energy loss due to gravitational radiation $t_{GW}$ 
is much longer than the orbital period $P$ as 
\begin{equation}
{t_{GW} \over P} \sim 15 \biggl({dc^2 \over 10GM}\biggr)^{5/2}
\biggl({M \over 4\mu}\biggr), \label{time}
\end{equation}
where $\mu$ and $d$ are the reduced mass and the separation of BNS's. 
Thus, BNS's adiabatically evolve radiating gravitational waves. 
However, when the orbital separation becomes $6-10GM/c^2$, 
they cannot maintain the circular orbit because of instabilities 
due to the GR gravity\cite{kidder} or the tidal field\cite{lai}. 
As a result of such instabilities, the circular orbit of BNS's 
changes into the plunging orbit to merge. 
This means that the nature of the signal of gravitational waves 
changes around the transition between the circular orbit and plunging 
one. Gravitational waves emitted at this transition region may 
bring us an important information about the structure of NS's 
because the location where the instability occurs will 
depend on the equation of state(EOS) of NS sensitively\cite{lai,joan}. 
Thus, it is very important to investigate the location of 
the innermost stable circular orbit(ISCO) of BNS's. 

As mentioned above, the ISCO is determined not only by the GR effects, 
but also by the hydrodynamic one. We emphasize that the tidal effects 
depend strongly on the structure of NS. Here, NS is a GR object 
because of its compactness, $Gm/c^2R\sim 0.2$, where $m$ and $R$ 
are the mass and radius of NS. 
Thus, in order to know the location of the ISCO accurately, 
we need to solve the GR hydrodynamic equations in general.  
A strategy to search the ISCO in GR manner is as follows; 
since the timescale of the energy loss is much longer than 
the orbital period according to Eq.$(\ref{time})$, we may suppose that 
the motion of BNS's is composed of 
the stationary part and the small radiation reaction part. From 
this physical point of view, we may consider that BNS's evolve 
quasi-stationally, and we can take the following procedure; 
first, neglecting the evolution due to gravitational radiation, 
equilibrium configurations are constructed, and then 
the radiation reaction is taken into account as a correction 
to the equilibrium configurations. 
The ISCO is determined from the point, where the dynamical instability 
for the equilibrium configurations occurs. 
It may be a grand challenge, however, to distinguish the stationary part 
from the nonstationary one in general relativity. 
As Detweiler has pointed out\cite{detweiler}, a stationary solution of 
the Einstein equation with standing gravitational waves, which will be 
constructed by adding the incoming waves from infinity, may be a valuable 
approximation to physically realistic solutions. 
However, these solutions are not asymptotically flat\cite{detweiler} 
because GWs contribute to the total energy of the system and the total 
energy of GWs inside a radius $r$ grows linearly with $r$. 
The lack of asymptotic flatness forces us to consider 
only a bounded space and impose boundary conditions in the near zone.
Careful consideration will be necessary to find out an appropriate boundary 
condition for describing the physically realistic system in the near zone. 

Recently, Wilson and his collaborators\cite{wilson} proposed a 
simirelativistic approximation method in order to calculate the 
equilibrium configuration of BNS's just before merging. In their method, 
they assume the line element as 
\begin{equation}
ds^2=-(\alpha^2-\beta_i \beta^i)c^2 dt^2+2 \beta_i c dt dx^i+\psi^4 dx^3,
\end{equation}
i.e., three metric $\gamma_{ij}$ is chosen as the conformal 
flat(i.e., $\gamma_{ij}=\psi^4\delta_{ij}$), and solve only the 
constraint equations in the Einstein equation. In their approach, 
they claim that they ignore only the contribution of gravitational waves, 
but it is not correct at all; as shown in previous post-Newtonian(PN) 
analyses\cite{shafer,asada,rieth}, 
the tensor potential term exists in the three metric 
even if we ignore the radiation reaction of gravitational waves(i.e., 
$\psi^{-4}\gamma_{ij}\not=\delta_{ij}$). Since such a term appears 
from the second PN order in the PN approximation, 
the accuracy of their results is less than the 2PN order: 
In reality, from results by Cook et al.\cite{cook} in which they obtain 
equilibrium configurations of the axisymmetric NS using both 
the Einstein equation and Wilson's method, 
we can see that some quantities obtained from Wilson's scheme, such as 
the lapse function, the three metric, the angular velocity, and so on,  
deviate from the exact solution by about $O((Gm/Rc^2)^2)$. 
This seems to indicate that their approach for the system of BNS's 
is valid only at the 1PN level from the PN point of view. 
Furthermore, the meaning of their approximation is obscure:  
It is not clear at all how to estimate errors due to such an approximation 
scheme and in which situation but the spherical symmetric system, the 
scheme based on the assumption of the conformal flatness is justified. 

In contrast with Wilson's method, the meaning of the PN approximation is 
fairly clear: In the PN approximation, the metric is formally expanded with 
respect to $c^{-1}$ assuming the slow motion and weak self-gravity of matter. 
If we will take into account the next PN order, the accuracy of 
approximate solutions will be improved. This means that we can estimate 
the order of magnitude of the error due to the ignorance of higher PN terms. 
Also, in the PN approximation, we can distinguish the radiation reaction 
terms, which begin at the 2.5PN order\cite{esposito}, 
from other terms in the metric. 
Thus, it is possible to construct the equilibrium configuration of BNS's 
without the radiation reaction terms in the 2PN approximation. 

We schematically describe two approaches in Tables 1(a) and 1(b). 
As mentioned above, in close binary of NS's, it is important to take into 
account GR effects to orbital motion  as well as to the internal structure 
of each NS. 
As for the orbital motion, there exist two parameters; 
one is the PN parameter $v/c$ and the other is the mass ratio $\eta$ of 
the reduced mass $\mu$ to the total mass $M$, and both parameters 
are less than unity. 
Thus, the physical quantities  such as the orbital frequency are expanded 
with respect to them. 
In Table 1(a), we show schematically various levels of approximations 
in terms of $v/c$ and $\eta$. 
If all terms in a level are taken into account in the 2PN approximation, 
we mark $P^2N$, while $W$ means that all terms in the marked level are 
taken into account in Wilson's approach. 
From Table 1(a), we see that the 2PN approximation can include all 
corrections in $\eta$ up to the 2PN order in contrast with Wilson's approach. 
On the other hand, Wilson's approach will hold completely in the test 
particle limit, i.e., at $O(\eta^0)$, whereas even in this limit 
the 2PN approximation is not valid at higher PN orders. 
As for the internal structure of each NS, there also exist two small 
parameters; one is the compactness $Gm/c^2R$ and the other is the 
deformation parameter from its spherical shape, such as an ellipticity $e$. 
In this case, the PN approximation becomes an expansion in terms of $Gm/c^2R$. 
In Table 1(b), we also show various levels of approximation in terms of 
these parameters. 
Although Wilson's approach is exact for spherical NS's, it is not valid 
in nonspherical cases even at the 2PN order. 
On the other hand, in the 2PN approximation, the spherical compact star 
cannot be obtained correctly in contrast with Wilson's approach. 
In this way, the 2PN approximation has a week point: 
Although it can take into account all effects up to the 2PN order, 
it is inferior to Wilson's approach when we take a test-particle limit, 
$\eta\to 0$, or we describe an exactly spherical NS. 
However, as shown below, the error due to the ignorance of higher PN terms 
in those cases is not so large . 

To estimate the error due to the ignorance of the higher PN terms, let us 
compare the GR exact solutions with their PN approximations. 
First, we consider a small star of mass $\mu$ orbiting 
a Schwarzschild black hole of 
mass $m_{bh} \gg \mu$. In this case, we may consider that the small star 
moves on the geodesic around the Schwarzschild black hole, and 
the orbital angular velocity becomes\cite{kidder}
\begin{equation}
\Omega=\sqrt{ {Gm_{bh} \over (\bar r+Gm_{bh}c^{-2})^3 } },\label{bheq}
\end{equation}
where $\bar r$ is the coordinate radius of the orbit in the harmonic 
coordinate. In the PN approximation, Eq.$(\ref{bheq})$ becomes
\begin{equation}
\Omega=\sqrt{ {Gm_{bh} \over \bar r^3} }\biggl\{1- {3Gm_{bh} \over 
2\bar r c^2}+{15 \over 8}
\Bigl({Gm_{bh} \over \bar r c^2}\Bigr)^2+O(c^{-6}) \biggr\}. 
\label{bhbheq}
\end{equation}
Comparing Eq.$(\ref{bheq})$ with Eq.$(\ref{bhbheq})$, it is found 
that the error size of the 
2PN angular velocity is $\sim 0.3\%$ at $\bar r=9Gm_{bh}c^{-2}$, and 
$\sim 1\%$  at $\bar r=6Gm_{bh}c^{-2}$. Thus, the 2PN approximation seems 
fairly good to describe the motion of relativistic binary stars just 
before coalescence. 
Next, we consider a spherical NS of a uniform density 
in order to investigate the applicability of the PN approximation for 
determination of the internal structure of NS's. 
In this model, the pressure, $P$, and the density, $\rho=$const., 
are related with each other\cite{compact}: 
\begin{eqnarray}
{P \over \rho c^2}&=&{ (1-2Gmr_s^2/c^2R^3)^{1/2}-(1-2Gm/c^2R)^{1/2} 
\over 3(1-2Gm/c^2R)^{1/2}-(1-2Gmr_s^2/c^2R^3)^{1/2} } \nonumber\\
&=&{1 \over 2}{Gm \over c^2R}\Bigl(1-{r_s^2 \over R^2}\Bigr)
+{G^2 m^2 \over c^4 R^2}\Bigl(1-{r_s^2 \over R^2}\Bigr)
+{G^3 m^3 \over c^6 R^3}\Bigl({17 \over 8}-{19 r_s^2 \over 8 R^2}
+{3 r_s^4 \over 8 R^4}-{r_s^6 \over 8 R^6}\Bigr) 
+O(c^{-8}), \label{ppppeq}
\end{eqnarray}
where $r_s$ is the coordinate radius in the Schwarzschild coordinate 
and terms of order $c^{-2}$, $c^{-4}$ and $c^{-6}$ denote 
Newtonian, 1PN and 2PN terms respectively. 
In the second line in Eq.$(\ref{ppppeq})$, we expand the equation in power 
of $Gm/c^2R$ regarding it as a small quantity. 
In fig.1, we shows the error, $1-\tilde P/P$, in Newtonian, 
1PN and 2PN cases 
as a function of $r_s$ for $R=5Gm/c^2$(solid lines) and 
$8Gm/c^2$(dotted lines), where $\tilde P$ 
denotes the PN approximate pressure. 
It is found that the discrepancy in the Newtonian treatment is very large, 
while in the 2PN approximation the error is less than 10$\%$. 
In this way, we can estimate rigidly the typical error size 
in the 2PN approximation. 
Furthermore, the accuracy is fairly good if the NS is 
not extremely compact; the 2PN approximation will be fairly accurate 
if the radius of NS is larger than $\sim$10km. 

Thus, in the present paper, we develop a formalism to obtain equilibrium 
configurations of uniformly rotating fluid in the 2PN order as a first step.
In section 2, we review the basic equations up to the 2PN order. 
In section 3, we rewrite the Poisson equation for potential functions, 
which are described in section 2, 
into useful forms in which the source terms of the Poisson equations 
decrease rapidly enough($O(r^{-4})$). 
In section 4, we show a formulation to obtain numerically equilibrium 
solutions of uniformly rotating fluid in the 2PN approximation: 
Taking into account the formulation in the first PN 
approximation\cite{shibata}, 
we further rewrite potentials defined in section 3
into a polynomial form in the 
angular velocity, $\Omega$. Then, we transform the integrated 
Euler equation into the polynomial form in $\Omega^2$ 
so that the convergence property in iteration procedures may be much 
improved.
For the sake of analysis for numerical results, we describe the 
2PN expression of the conserved quantities, such as the conserved mass, 
the ADM mass, the total energy and the total angular momentum in section 5.
Section 6 is devoted to summary. 
Throughout this paper, $G$ and $c$ denote the gravitational constant and the 
speed of light. Hereafter, we use units of $G=1$. 

\section{Formulation} 

We write the line element in the following form; 
\begin{equation}
ds^2=-(\alpha^2-\beta_i \beta^i)c^2dt^2+2\beta_i c dt dx^i+
\psi^4 \tilde \gamma_{ij}dx^i dx^j, 
\end{equation}
where we define ${\rm det}(\tilde \gamma_{ij})=1$. 
To fix the gauge condition in the time coordinate, 
we use the maximal slice condition $K_i^{~i}=0$, 
where $K_i^{~i}$ is the trace part of the extrinsic curvature, $K_{ij}$. 
As the spatial gauge condition, we adopt 
the transverse gauge $\tilde \gamma_{ij,j}=0$ in order to remove the 
gauge modes from $\tilde \gamma_{ij}$. 
In this case, up to the 2 PN approximation, each metric variable 
is expanded as\cite{asada}
\begin{eqnarray}
\psi&=&1+{1 \over c^2}{U \over 2}+{1 \over c^4}\four\psi+O(c^{-6}),\\
\alpha&=&1-{1 \over c^2}U+{1 \over c^4}\Bigl({U^2 \over 2}+X\Bigr)
+{1 \over c^6}\six\alpha+O(c^{-7}), \\
\beta^i&=&{1 \over c^3}\three\beta_i+{1 \over c^5}\five\beta_i+
O(c^{-7}),\\
\tilde \gamma_{ij}&=&\delta_{ij}+{1 \over c^4}h_{ij}+O(c^{-5}).
\end{eqnarray}

As for the energy-momentum tensor of the Einstein equation, 
we consider the perfect fluid as 
\begin{equation}
T_{\mu\nu}=\Bigl(\rho c^2+\rho \varepsilon+P \Bigr)u_{\mu}u_{\nu}
+P g_{\mu\nu}. 
\end{equation}
For simplicity, we assume that the 
matter obeys the polytropic equation of state(EOS);
\begin{equation}
P=(\Gamma-1)\rho \varepsilon=K \rho^{\Gamma}, 
\end{equation}
where $\Gamma$ and $K$ are the polytropic exponent and polytropic constant, 
respectively. Up to the 2PN order, 
the four velocity is expanded as\cite{chandra,asada}
\begin{eqnarray}
u^0&=&1+{1 \over c^2}\Bigl({1 \over 2}v^2+U\Bigr)
+{1 \over c^4}\Bigl({3 \over 8}v^4+{5 \over 2}v^2U
+{1 \over 2}U^2+\three\beta_i v^i-X \Bigr)+O(c^{-6}),\nonumber\\
u_0&=&-\biggl[ 1+{1 \over c^2}\Bigl({1 \over 2}v^2-U\Bigr)
+{1 \over c^4}\Bigl({3 \over 8}v^4+{3 \over 2}v^2U
+{1 \over 2}U^2+X \Bigr)\biggr]+O(c^{-6}),\nonumber\\
u^i&=&{v^i \over c}
\biggl[1+{1 \over c^2}\Bigl({1 \over 2}v^2+U\Bigr)
+{1 \over c^4}\Bigl({3 \over 8}v^4+{5 \over 2}v^2U
+{1 \over 2}U^2+\three\beta_i v^i-X \Bigr)\biggr]+O(c^{-7}),\nonumber\\
u_i&=&{v^i \over c}+{1 \over c^3}\Bigl\{\three\beta_i
+v^i\Bigl({1 \over 2}v^2+3 U\Bigr)\Bigr\} 
+{1 \over c^5}\Bigl\{\five\beta_i+\three\beta_i\Bigl({1 \over 2}v^2+3U\Bigr)
+h_{ij}v^j \nonumber\\
&&~~~~~~~~~~~~~~+v^i\Bigl({3 \over 8}v^4+{7 \over 2}v^2 U+4U^2-X+4\four\psi 
+\three\beta_j v^j\Bigr)\Bigr\}+O(c^{-6}), 
\label{uueq}
\end{eqnarray}
where $v^i=u^i/u^0$ and $v^2=v^i v^i$. Since we need $u^0$ up to 3PN 
order to obtain the 2PN equations of motion, we derive it here. 
Using Eq.$(\ref{uueq})$, we can calculate $(\alpha u^0)^2$ up to 3PN order as 
\begin{eqnarray}
(\alpha u^0)^2&=&1 + \psi^{-4}\tilde \gamma^{ij}u_i u_j \nonumber\\
&=&1+{v^2 \over c^2}+{1 \over c^4}\Bigl(2\three\beta_j v^j+4Uv^2+v^4\Bigr)
+{1 \over c^6}\Bigl\{\three\beta_j\three\beta_j+8\three\beta_j v^j U 
+h_{ij}v^i v^j \nonumber\\
&&~~~~~~~~~+2\five\beta_iv^i
+\Bigl(4\three\beta_j v^j+4\four\psi+{15 \over 2}U^2-2X\Bigr)v^2
+8Uv^4+v^6\Bigr\}+O(c^{-7}),
\end{eqnarray}
where we use $\tilde \gamma^{ij}=\delta_{ij}-c^{-4}h_{ij}+O(c^{-5})$. 
Thus, we obtain $u^0$ up to the 3PN order as 
\begin{eqnarray}
u^0&=&1+{1 \over c^2}\Bigl({1 \over 2}v^2+U\Bigr)
+{1 \over c^4}\Bigl({3 \over 8}v^4+{5 \over 2}v^2U
+{1 \over 2}U^2+\three\beta_i v^i-X \Bigr) \nonumber\\
&&+{1 \over c^6}\Bigl\{-\six\alpha
+{1 \over 2}\Bigl( \three\beta_j\three\beta_j+h_{ij}v^iv^j\Bigr)
+\five \beta_j v^j + 5\three\beta_j v^j U -2UX \nonumber\\
&&~~~~~+\Bigl({3 \over 2}
\three\beta_j v^j+2\four\psi+6U^2-{3 \over 2} X\Bigr)v^2 
+{27 \over 8}Uv^4+{5 \over 16}v^6\Bigr\}+O(c^{-7}). \label{utime}
\end{eqnarray}

Substituting PN expansions of metric and matter variables into the 
Einstein equation, and using the polytropic EOS, 
we find that the metric variables obey the following
Poisson equations\cite{asada};
\begin{eqnarray}
&&\Delta U=-4\pi \rho, \\
&&\Delta X=4\pi\rho\Bigl(2 v^2+2U+(3\Gamma-2)\varepsilon
\Bigr), \\
&&\Delta \four\psi=-2\pi\rho\Bigl(v^2+\varepsilon+{5 \over 2}U \Bigr), \\
&&\Delta \three\beta_i=16\pi \rho v^i-\dot U_{,i}, \\
&&\Delta \five\beta_i =16\pi\rho\biggl[v^i \Bigl(v^2+2U+\Gamma\varepsilon
\Bigr)+\three\beta_i\biggr]-4U_{,j}
\Bigl(\three\beta_{i,j}+\three\beta_{j,i}
-{2 \over 3}\delta_{ij}\three\beta_{k,k}\Bigr) \nonumber\\
&&{\hskip 2cm} -2\four\dot\psi_{,i}
      +{1 \over 2}(U \dot U)_{,i}+(\three\beta_l U_{,l})_{,i}, \\
&&\Delta  h_{ij}=\Bigl(U U_{,ij}-{1 \over 3}\delta_{ij}U \Delta_{flat}U
-3U_{,i} U_{,j}+\delta_{ij}U_{,k} U_{,k} \Bigr)
-16\pi \Bigl(\rho v^i v^j -{1 \over 3}\delta_{ij}\rho v^2 \Bigr) \nonumber\\
&&\hskip 1cm  -\Bigl(\three\dot\beta_{i,j}+\three\dot\beta_{j,i}
-{2 \over 3}\delta_{ij}\three\dot\beta_{k,k} \Bigr) 
-2\Bigl( (X+2\four\psi)_{,ij}-{1 \over 3}\delta_{ij}
\Delta (X+2\four\psi) \Bigr), \\
&&\Delta \six\alpha=4\pi\rho\biggl[2v^4
+2v^2\Bigl(5U+\Gamma\varepsilon\Bigr)+(3\Gamma-2)\varepsilon U
+4\four\psi+X+4\three\beta_i v^i \biggr] \nonumber\\
&&{\hskip 3cm}-h_{ij}U_{,ij}
-{3 \over 2}U U_{,l}U_{,l}+U_{,l}(2 \four\psi -X)_{,l} \nonumber\\
&&{\hskip 3cm} +{1 \over 2}
\three\beta_{i,j}\Bigl(\three\beta_{i,j}+\three\beta_{j,i}
-{2 \over 3}\delta_{ij}\three\beta_{k,k}\Bigr), 
\end{eqnarray}
where $\Delta$ is the flat Laplacian, and $~\cdot~$ denotes $\pa/\pa t$. 

Equations of motion for fluid are derived from 
\begin{equation}
\nabla_{\mu} T^{\mu}_{~\nu}=0. \label{eom}
\end{equation}
In this paper, we consider the uniformly rotating fluid around $z$-axis 
with the angular velocity $\Omega$, i.e., 
\begin{equation}
v^i=\epsilon_{ijk}\Omega^j x^k=(-y\Omega, x\Omega, 0), 
\end{equation}
where we choose $\Omega^j=(0,0,\Omega)$ and $\epsilon_{ijk}$ is the completely 
anti-symmetric unit tensor. 
In this case, the following relations hold;
\begin{equation}
\Bigl({\partial \over \partial t}+\Omega{\partial \over \partial \varphi}
\Bigr) Q
=\Bigl({\partial \over \partial t}+\Omega{\partial \over \partial \varphi}
\Bigr) Q_i
=\Bigl({\partial \over \partial t}+\Omega{\partial \over \partial \varphi} 
\Bigr)Q_{ij}=0,\label{killeq}
\end{equation} 
where $Q$, $Q_i$ and $Q_{ij}$ are arbitrary scalars, vectors, and tensors, 
respectively. Then, Eq.$(\ref{eom})$ can be integrated as\cite{lightman}  
\begin{equation}
\int {dP \over \rho c^2+\rho\varepsilon+P}=\ln u^0+C,\label{euler}
\end{equation}
where $C$ is a constant. For the polytropic EOS, Eq.$(\ref{euler})$ becomes 
\begin{equation}
\ln \biggl[1+{\Gamma K \over c^2(\Gamma-1)} \rho^{\Gamma-1} \biggr]
= \ln u^0+C, \label{eulerf}
\end{equation}
or
\begin{equation}
1+{\Gamma K \over c^2(\Gamma-1)} \rho^{\Gamma-1}=u^0 \exp(C). 
\label{euleri}
\end{equation}
Using Eq.$(\ref{utime})$, the 2PN approximation of Eq.$(\ref{eulerf})$ 
is written as
\begin{eqnarray}
H-{H^2 \over 2c^2}+{H^3 \over 3c^4}&=&{v^2 \over 2}+U
+{1 \over c^2}\biggl(2Uv^2+{v^4 \over 4}-X+\three\beta_i v^i \biggl) 
\nonumber\\
&&+{1 \over c^4}\biggl(-\six\alpha +{1 \over 2}\three\beta_i\three\beta_i
+4\three\beta_i v^i U-{U^3 \over 6}+\three\beta_i v^i v^2+2 \four \psi v^2 
\nonumber\\
&&{\hskip 1.2cm}
+{15\over 4} U^2 v^2+2Uv^4 +{1 \over 6} v^6-UX-v^2 X+\five\beta_i v^i
+{1 \over 2}h_{ij}v^iv^j\biggr)+C, \label{bern}
\end{eqnarray}
where $H=\Gamma K \rho^{\Gamma-1}/(\Gamma-1)$, $v^2=R^2\Omega^2$ and 
$R^2=x^2+y^2$. Note that Eq.$(\ref{bern})$ can be also obtained 
from the 2PN Euler equation like in the first PN case\cite{jcb,shibata}. 
If we solve the coupled equations (2.11-17) and $(\ref{bern})$, 
we can obtain equilibrium configurations of the non-axisymmetric 
uniformly rotating body. 

\section{Derivation of the Poisson equation of compact sources for 
\lowercase{$h_{ij}$, $\three\beta_i$ and $\five\beta_i$}}

In section 2, we derive the Poisson equations for metric variables. 
However, the source terms in the Poisson equations for $\three \beta_i$, 
$\five \beta_i$, and $h_{ij}$ fall off slowly as $r \rightarrow \infty$ 
because these terms behave as $O(r^{-3})$ at $r \rightarrow \infty$. 
These Poisson equations do not take convenient forms when we try to solve 
them as the boundary value problem in numerical calculation. 
Hence in the following, we rewrite them into other convenient forms 
in numerical calculation. 

As for $h_{ij}$, first of all, we split the equation into three parts 
as\cite{asada}
\begin{eqnarray}
\Delta h^{(U)}_{ij}&=&U\Bigl(U_{,ij}-{1 \over 3}\delta_{ij}\Delta U\Bigr)
-3U_{,i}U_{,j}+\delta_{ij}U_{,k}U_{,k} \equiv -4\pi S^{(U)}_{ij},   \\
\Delta h^{(S)}_{ij}&=&
-16\pi\Bigl(\rho v^iv^j-{1 \over 3}\delta_{ij}\rho v^2\Bigr),   \\
\Delta h^{(G)}_{ij}&=&-\Bigl(\three\dot\beta_{i,j}+\three\dot\beta_{j,i}
-{2 \over 3}\delta_{ij}\three\dot\beta_{k,k} \Bigr) \nonumber\\
&&{\hskip 1.6cm}-2\Bigl( (X+2\four\psi)_{,ij}-{1 \over 3}\delta_{ij}
\Delta (X+2\four\psi) \Bigr). \label{hijG} 
\end{eqnarray}
The equation for $h^{(S)}_{ij}$ has a compact source, and also 
the source term of $h^{(U)}_{ij}$ behaves as $O(r^{-6})$ at 
$r \rightarrow \infty$, so that Poisson equations for them are 
solved easily as the boundary value problem. 
On the other hand, the source term of $h^{(G)}_{ij}$ behaves 
as $O(r^{-3})$ at $r \rightarrow \infty$, 
so that it seems troublesome to solve the equation for it 
as the boundary value problem. In order to solve the equation for 
$h^{(G)}_{ij}$ as the boundary value problem, we had better 
rewrite the equation into useful forms. 
As shown in a previous paper\cite{asada}, 
Eq.$(\ref{hijG})$ is integrated to give 
\begin{eqnarray}
h_{ij}^{(G)}&=&
2{\pa \over \pa x^i}\int(\rho v^j)^{\cdot}\vert {\bf x}-{\bf y} \vert d^3y
+2{\pa \over \pa x^j}\int(\rho v^i)^{\cdot}\vert {\bf x}-{\bf y} \vert d^3y
+\delta_{ij}\int\ddot\rho \vert {\bf x}-{\bf y} \vert d^3y \nonumber\\
&&+{1 \over 12}{\pa^2 \over \pa x^i \pa x^j}\int\ddot\rho 
\vert {\bf x}-{\bf y} \vert^3 d^3y
+{\pa^2 \over \pa x^i \pa x^j}\int\Bigl(\rho v^2+3P-{\rho U \over 2} \Bigr)
\vert {\bf x}-{\bf y} \vert d^3y \nonumber\\
&&~~~~~~~~~~~~~~-{2 \over 3}\delta_{ij}\int{\Bigl(\rho v^2+3P-\rho U / 2 \Bigr)
\over \vert {\bf x}-{\bf y} \vert} d^3y. \label{haeq}
\end{eqnarray}
Using the relations
\begin{eqnarray}
&&\ddot \rho=-(\rho v^j)_{,j}^{\cdot}+O(c^{-2}),\nonumber\\
&&\dot v^i=0,\nonumber\\
&&v^i x^i=0, 
\end{eqnarray}
Eq.$(\ref{haeq})$ is rewritten as
\begin{eqnarray}
h_{ij}^{(G)}&=&
{7 \over 4}\biggl[\int(\rho v^j)^{\cdot}
{x^i-y^i \over |{\bf x}-{\bf y}|} d^3y
+\int(\rho v^i)^{\cdot}{x^j-y^j \over |{\bf x}-{\bf y}|} d^3y\biggr]
-\delta_{ij}x^k
\int {(\rho v^k)^{\cdot} \over |{\bf x}-{\bf y}|}d^3y \nonumber\\
&&-{1 \over 8}x^k \biggl[ {\partial \over \partial x^i}\int (\rho v^k)^{\cdot}
{x^j-y^j \over |{\bf x}-{\bf y}|} d^3y
+{\partial \over \partial x^j}\int (\rho v^k)^{\cdot}
{x^i-y^i \over |{\bf x}-{\bf y}|} d^3y \biggr] \nonumber\\
&&+{1 \over 2}\biggl[
{\partial \over \partial x^i}\int\Bigl(\rho v^2+3P-{\rho U \over 2} \Bigr)
{x^j-y^j \over |{\bf x}-{\bf y}|}d^3y
+{\partial \over \partial x^j}\int\Bigl(\rho v^2+3P-{\rho U \over 2} \Bigr)
{x^i-y^i \over |{\bf x}-{\bf y}|}d^3y \biggr] \nonumber\\
&&{\hskip 3cm}-{2 \over 3}\delta_{ij}\int{\Bigl(\rho v^2+3P-\rho U / 2 \Bigr)
\over \vert {\bf x}-{\bf y} \vert} d^3y. \label{hbeq}
\end{eqnarray}
From Eq.$(\ref{hbeq})$, it is found that $h_{ij}^{(G)}$ is written as
\begin{eqnarray}
 h_{ij}^{(G)}&=&{7 \over 4}
\Bigl(x^i \three \dot P_j+x^j \three \dot P_i-\dot Q^{(T)}_{ij}
-\dot Q^{(T)}_{ji}\Bigr)
-\delta_{ij}x^k \three \dot P_k \nonumber\\
&&-{1 \over 8} x^k 
\biggl[{\partial \over \partial x^i}\Bigl(x^j \three \dot P_k
-\dot Q^{(T)}_{kj} \Bigr)
+{\partial \over \partial x^j}\Bigl(x^i \three \dot P_k
-\dot Q^{(T)}_{ki} \Bigr)\biggr] \nonumber\\
&&+{1 \over 2}
\biggl[{\partial \over \partial x^i}\Bigl(x^j Q^{(I)}- Q^{(I)}_j \Bigr)
+{\partial \over \partial x^j}\Bigl(x^i Q^{(I)}- Q^{(I)}_i \Bigr) \biggr]
-{2 \over 3}\delta_{ij} Q^{(I)}, 
\end{eqnarray}
where 
\begin{eqnarray}
&&\Delta \three P_i=-4\pi \rho v^i, \\
&&\Delta Q^{(T)}_{ij}=-4\pi \rho v^i x^j, \\
&&\Delta Q^{(I)}=-4\pi \Bigl(\rho v^2+3P-{1 \over 2}\rho U\Bigr), \\
&&\Delta Q^{(I)}_i
     =-4\pi \Bigl(\rho v^2+3P-{1 \over 2}\rho U \Bigr)x^i. 
\end{eqnarray}
Therefore, $h_{ij}^{(G)}$ can be deduced from variables which satisfy 
the Poisson equations with compact sources. 

The source terms in the Poisson equations for $\three\beta_i$ and 
$\five\beta_i$ also fall off slowly. 
However, if we rewrite them as\cite{asada} 
\begin{eqnarray}
\three\beta_i&=&-4\three P_i-{1 \over 2}\Bigl(x^i \dot U-\dot q_i\Bigr), \\
\five \beta_i&=&-4\five  P_i-{1 \over 2}\Bigl(2x^i \four \dot \psi
-\dot \eta_i \Bigr), 
\end{eqnarray}
where
\begin{eqnarray}
&&\Delta q_i=-4\pi\rho x^i , \\
&&\Delta \five P_i=-4\pi \rho\biggl[v^i\Bigl(v^2+2U+\Gamma\varepsilon \Bigr)
+\three \beta_i\biggr]+U_{,j}\Bigl(\three\beta_{i,j}+\three\beta_{j,i}
-{2 \over 3}\delta_{ij}\three\beta_{k,k}\Bigr) \nonumber\\
&&{\hskip 4cm}-{1 \over 8}(\dot U U)_{,i}
-{1 \over 4}(\three \beta_l U_{,l})_{,i}, \\
&&\Delta \eta_i=-4\pi\rho\Bigl(v^2+\varepsilon+{5 \over 2}U\Bigr)x^i, 
\end{eqnarray}
then $\three\beta_i$ and $\five\beta_i$ can be obtained by solving 
the Poisson equations in which the fall-off of the source terms is 
fast enough, $O(r^{-5})$, for numerical calculation. 
Note that, using the relation $\three P_i=\epsilon_{izk}q_k \Omega$ 
and Eqs.$(\ref{killeq})$, $\three \beta_i$ and $\five\beta_i$ may be written as 
\begin{eqnarray}
&&\three \beta_i =
\Omega\Bigl\{-4\epsilon_{izk} q_k+{1 \over 2}\Bigl(x^i U_{,\varphi}
-q_{i,\varphi}\Bigr)\Bigr\}\equiv \Omega\three \hat \beta_i, \\
&&\five \beta_i =
\Omega\Bigl\{-4\five \hat P_i+{1 \over 2}\Bigl(2x^i \four\psi_{,\varphi}
-\eta_{i,\varphi}\Bigr)\Bigr\}, 
\end{eqnarray}
where
\begin{eqnarray}
\Delta \five \hat P_i &=&-4\pi\rho\biggl[ \epsilon_{izk}x^k\Bigl(v^2+2U
+\Gamma \varepsilon\Bigr)+\three \hat \beta_i\biggr]
+U_{,j}\Bigl(\three\hat \beta_{i,j}+\three\hat \beta_{j,i}
-{2 \over 3}\delta_{ij}\three\hat \beta_{k,k}\Bigr) \nonumber\\
&&{\hskip 4cm} +{1 \over 8}(UU_{,\varphi})_{,i}-{1 \over 4}
(\three\hat \beta_k U_{,k})_{,i}. 
\end{eqnarray}

\section{Derivation of basic equations}

In this section, we derive the basic equation 
which has a suitable form
to construct equilibrium configurations of uniformly rotating body in 
numerical calculation: 
Although equilibrium configurations 
can be formally obtained by solving Eq.$(\ref{bern})$ as well as
metric potentials, $U$, $X$, $\four\psi$, $\six\alpha$, 
$\three \beta_i$, $\five\beta_i$ and $h_{ij}$, 
they do not take convenient forms for numerical calculation. 
Thus, we here change Eq.$(\ref{bern})$ into other forms appropriate 
to obtain numerically equilibrium configurations. 

In numerical calculation, the standard method to obtain equilibrium 
configurations is as follows\cite{hachisu,oohara,shibata}; 

\noindent
(1) We give a trial density configuration for $\rho$. 

\noindent
(2)
We solve the Poisson equations. 

\noindent
(3)
Using Eq.$(\ref{bern})$, we give a new density configuration. 

\noindent
These procedures are repeated until a sufficient convergence is achieved. 
Here, at (3), we need to specify unknown constants, $\Omega$ and $C$. 
In standard numerical methods\cite{hachisu,oohara}, 
these are calculated during iteration fixing 
densities at two points; i.e., if we put $\rho_1$ and $\rho_2$ at $x_1$ and 
$x_2$ into Eq.$(\ref{bern})$, they become two simultaneous equations 
for $\Omega$ and $C$. Hence, we can calculate them. 
However, the procedure is not so simple in the PN case: $\Omega$ is 
included in the source of the Poisson equations for 
the variables such as $X$, $\four \psi$, $\six\alpha$, 
$\eta_i$, $\five \hat P_i$, $h_{ij}^{(S)}$, $Q^{(T)}_{ij}$, $Q^{(I)}$ and 
$Q^{(I)}_i$. 
Thus, if we use Eq.$(\ref{bern})$ as it is, equations for 
$\Omega$ and $C$ become implicit equations for $\Omega$. As found in 
a previous paper\cite{shibata}, in such a situation, the convergence 
to a solution is very slow. 
Therefore, we transform those equations into other forms 
in which the potentials as well as 
Eq.$(\ref{bern})$ become explicit polynomial equations in $\Omega$.

First of all, we define $q_2$, $q_{2i}$, $q_4$, $q_u$, $q_e$ and 
$q_{ij}$ which satisfy
\begin{eqnarray}
\Delta q_2 &=&-4\pi\rho R^2, \\
\Delta q_{2i} &=&-4\pi\rho R^2 x^i, \\
\Delta q_4 &=&-4\pi\rho R^4, \\
\Delta q_u &=&-4\pi\rho U, \\
\Delta q_e &=&-4\pi\rho \varepsilon, \\
\Delta q_{ij} &=&-4\pi\rho x^i x^j. 
\end{eqnarray}
Then, $X$, $\four\psi$, $Q^{(I)}$, $Q_i^{(I)}$, $\eta_i$, 
$\five \hat P_i$, $Q_{ij}^{(T)}$, and $h^{(S)}_{ij}$ are written as 
\begin{eqnarray}
X&=&-2q_2\Omega^2-2 q_u-(3\Gamma-2)q_e, \\
\four\psi&=&{1 \over 2}\Bigl(q_2\Omega^2+q_e+{5 \over 2}q_u\Bigr), \\
Q^{(I)}&=&q_2 \Omega^2+3(\Gamma-1)q_e-{1 \over 2}q_u 
\equiv q_2\Omega^2+Q^{(I)}_0, \\
Q_i^{(I)}&=&q_{2i}\Omega^2+Q_{0i}^{(I)}, \\
\eta_i&=&q_{2i}\Omega^2+\eta_{0i}, \\
\five \hat P_i&=&\epsilon_{izk}q_{2k}\Omega^2+\five P_{0i}, \\
Q_{ij}^{(T)}&=&\epsilon_{izl}q_{lj} \Omega, \\
h^{(S)}_{ij}&=&4\Omega^2\Bigl(\epsilon_{izk}\epsilon_{jzl}q_{kl}
-{1 \over 3}\delta_{ij}q_2\Bigr), 
\end{eqnarray}
where $Q_{0i}^{(I)}$, $\eta_{0i}$ and $\five P_{0i}$ satisfy
\begin{eqnarray}
\Delta Q_{0i}^{(I)} &=&-4\pi\Bigl( 3P- { 1 \over 2} \rho U\Bigr)x^i
=-4\pi\rho\Bigl( 3(\Gamma-1)\varepsilon-{1 \over 2}U \Bigr)x^i, \\
\Delta \eta_{0i} &=&-4\pi\rho\Bigl(\varepsilon+{5 \over 2}U\Bigr)x^i, \\
\Delta \five P_{0i} &=&-4\pi \rho \biggl[\epsilon_{izk} x^k \Bigl(2U
+\Gamma \varepsilon \Bigr)+\three \hat \beta_i\biggr]
+U_{,j} \Bigl( \three \hat \beta_{i,j}+
\three \hat \beta_{j,i}-{2 \over 3}\delta_{ij}\three\hat \beta_{k,k}\Bigr) 
\nonumber\\
&&{\hskip 3cm} +{1 \over 8}(UU_{,\varphi})_{,i}-{1 \over 4}
(\three \hat \beta_k U_{,k})_{,i} \equiv -4\pi S^{(P)}_i. 
\end{eqnarray}
Note that $\five \beta_i$ and $h_{ij}^{(G)}$ are the cubic and quadratic 
equations in $\Omega$, respectively, as
\begin{eqnarray}
\five \beta_i &=&\Omega\biggl[-4\five P_{0i}+{1 \over 2}\Bigl\{x^i
\Bigl(q_e+{5 \over 2}q_u\Bigr)_{,\varphi}-\eta_{0i,\varphi}\Bigr\}\biggr]
+\Omega^3\biggl[ -4\epsilon_{izk}q_{2k}
+{1 \over 2}\Bigl(x^iq_{2,\varphi}-q_{2i,\varphi}\Bigr)\biggr] \nonumber\\
&& \equiv \five \beta^{(A)}_i\Omega +\five \beta_i^{(B)}\Omega^3, \\
h_{ij}^{(G)} &=&{1 \over 2}\biggl[
{\pa \over \pa x^j}\Bigl(x^i Q^{(I)}_0-Q^{(I)}_{0i}\Bigr)+
{\pa \over \pa x^i}\Bigl(x^j Q^{(I)}_0-Q^{(I)}_{0j}\Bigr)
-{4 \over 3}\delta_{ij}Q^{(I)}_0\biggr] \nonumber\\
&&+\Omega^2 \biggl[{1 \over 2}\biggl\{
{\pa \over \pa x^j}\Bigl(x^i q_2-q_{2i}\Bigr)+
{\pa \over \pa x^i}\Bigl(x^j q_2-q_{2j}\Bigr)
-{4 \over 3}\delta_{ij}q_2 \biggr\} \nonumber\\
&&\hskip 2cm
-{7 \over 4}\Bigl(x^i\epsilon_{jzk}q_{k,\varphi}+x^j\epsilon_{izk}q_{k,\varphi}
-\epsilon_{izk}q_{kj,\varphi}-\epsilon_{jzk}q_{ki,\varphi}\Bigr)
+\delta_{ij}x^k \epsilon_{kzl} q_l \nonumber\\
&&\hskip 2cm
+{1 \over 8}x^k\biggl\{ {\pa \over \pa x^i}
\Bigl(x^j\epsilon_{kzl}q_{l,\varphi}-\epsilon_{kzl}q_{lj,\varphi}\Bigr)
+{\pa \over \pa x^j}
\Bigl(x^i\epsilon_{kzl}q_{l,\varphi}-\epsilon_{kzl}q_{li,\varphi}\Bigr)
\biggr\} \biggr] \nonumber\\
&& \equiv h_{ij}^{(A)}+h_{ij}^{(B)} \Omega^2 . 
\end{eqnarray} 
Finally, we write $\six\alpha$ as
\begin{equation}
\six\alpha=\six \alpha_0+\six \alpha_2 \Omega^2-2q_4\Omega^4, 
\end{equation}
where $\six \alpha_0$ and $\six \alpha_2$ satisfy
\begin{eqnarray}
\Delta \six \alpha_0&=&4\pi\rho\biggl[\Bigl(3\Gamma-2\Bigr) \varepsilon U
-\Bigl(3\Gamma-4\Bigr)q_e+3q_u\biggr] \nonumber\\
&&-\Bigl(h_{ij}^{(U)}+h_{ij}^{(A)}\Bigr)U_{,ij}-{3 \over 2}UU_{,l}U_{,l}
+U_{,l}{\pa \over \pa x^l}\Bigl({9 \over 2}q_u+(3\Gamma+1)q_e\Bigr) 
\nonumber\\
&\equiv& -4\pi S^{(\alpha_0)},  \\
\Delta \six \alpha_2&=&8\pi\rho R^2\Bigl(5U+\Gamma\varepsilon
+2\three\hat\beta_{\varphi} \Bigr)
-\Bigl(4\epsilon_{izk}\epsilon_{jzl}q_{kl}-{4 \over 3}\delta_{ij}q_2
+h_{ij}^{(B)}\Bigr)U_{,ij}+3q_{2,l}U_{,l} \nonumber\\
&& \hskip 2cm +{1 \over 2}
\three\hat\beta_{i,j} \Bigl(\three\hat\beta_{i,j}+\three\hat\beta_{j,i}
-{2 \over 3}\delta_{ij}\three\hat\beta_{k,k}\Bigr) \nonumber\\
&\equiv& -4\pi S^{(\alpha_2)} . 
\end{eqnarray}
Using the above quantities, Eq.$(\ref{bern})$ is rewritten as
\begin{equation}
H-{H^2 \over 2c^2}+{H^3 \over 3c^4}=A+B\Omega^2+D\Omega^4
+{R^6 \over 6c^4}\Omega^6+C, \label{berneq}
\end{equation}
where
\begin{eqnarray}
A&=&U+{1 \over c^2}\Bigl(2q_u+(3\Gamma-2)q_e\Bigr)+{1 \over c^4}\Bigl\{
-\six\alpha_0-{U^3 \over 6}+U\Bigl(2q_u+(3\Gamma-2)q_e\Bigr)\Bigr\},\nonumber\\
B&=&{R^2 \over 2}+{1 \over c^2}\Bigl(2R^2U+2q_2+\three\hat\beta_{\varphi}\Bigr)
+{1 \over c^4}\Bigl\{
-\six\alpha_2+{1 \over 2}\three\hat\beta_i\three\hat\beta_i
+4\three\hat\beta_{\varphi}U \nonumber\\
&&+(3\Gamma-1)q_eR^2+{9 \over 2}q_uR^2+{15 \over 4}U^2R^2+2q_2U+
\five\beta_{\varphi}^{(A)}
+{1 \over 2}\Bigl(h_{\varphi\varphi}^{(U)}+h_{\varphi\varphi}^{(A)}\Bigr)
\Bigr\}, \nonumber\\
D&=&{R^4 \over 4c^2}+{1 \over c^4}\Bigl\{
2q_4+\three\hat\beta_{\varphi}R^2+{7 \over 3}q_2R^2+2UR^4+
\five\beta_{\varphi}^{(B)}+{1 \over 2}\Bigl(h_{\varphi\varphi}^{(B)}
+4R^2q_{RR}\Bigr)\Bigr\}. 
\end{eqnarray}
Note that in the above, we use the following relations which hold 
for arbitrary vector $Q_i$ and symmetric tensor $Q_{ij}$, 
\begin{eqnarray}
Q_{\varphi}&=&-yQ_{x}+xQ_{y}, \nonumber\\ 
Q_{\varphi\varphi}&=&y^2Q_{xx}-2xyQ_{xy}+x^2Q_{yy}, \nonumber\\
R^2Q_{RR}&=&x^2Q_{xx}+2xyQ_{xy}+y^2Q_{yy}. 
\end{eqnarray}
We also 
note that source terms of Poisson equations for variables which appear in 
$A$, $B$ and $D$ do not depend on $\Omega$ 
explicitly. Thus, Eq.$(\ref{berneq})$ takes the desired form 
for numerical calculation.

In this formalism, we need to solve 29 Poisson equations for 
$U$, $q_x$, $q_y$, $q_z$, $\five P_{0x}$, $\five P_{0y}$, 
$\eta_{0x}$, $\eta_{0y}$, $Q^{(I)}_{0x}$, $Q^{(I)}_{0y}$, $Q^{(I)}_{0z}$, 
$q_2$, $q_{2x}$, $q_{2y}$, $q_{2z}$, $q_u$, $q_e$, 
$h_{xx}^{(U)}$, $h_{xy}^{(U)}$, $h_{xz}^{(U)}$, $h_{yy}^{(U)}$, $h_{yz}^{(U)}$,
$q_{xx}$, $q_{xy}$, $q_{xz}$, $q_{yz}$, $\six\alpha_0$, 
$\six\alpha_2$ and $q_4$. In Table 2, we show the list of the Poisson equations
to be solved. In Table 3, 
we also summarize what variables are needed to calculate 
the metric variables $U$, $X$, $\four\psi$, $\six \alpha$, $\three \beta_i$, 
$\five\beta_i$, $h_{ij}^{(U)}$,  $h_{ij}^{(S)}$, $h_{ij}^{(A)}$ and 
$h_{ij}^{(B)}$. Note that we do not need $\five P_{0z}$, 
$\eta_{0z}$, and $q_{zz}$ because they do not appear in any equation. 
Also, we do not have to solve the Poisson 
equations for $h^{(U)}_{zz}$ and $q_{yy}$ because they can be calculated from 
$h^{(U)}_{zz}=-h^{(U)}_{xx}-h^{(U)}_{yy}$ and $q_{yy}=q_2-q_{xx}$. 

In order to derive $U$, $q_i$, $q_2$, $q_{2i}$, $q_4$, $q_e$ and $q_{ij}$, 
we do not need any other potential because 
only matter variables appear in the source terms of their 
Poisson equations. On the other hand, 
for $q_u$, $Q_{0i}^{(I)}$, $\eta_{0i}$ and $h_{ij}^{(U)}$, 
we need the Newtonian potential $U$, and 
for $\five P_{0i}$, $\six\alpha_0$ and $\six\alpha_2$, 
we need the Newtonian as well as PN potentials. 
Thus, $U$, $q_i$, $q_2$, $q_{2i}$, $q_4$, $q_e$ and $q_{ij}$ 
must be solved first, and then $q_u$, $Q_{0i}^{(I)}$, $\eta_{0i}$, 
$h_{ij}^{(U)}$, $\five P_{0i}$ and $\six\alpha_2$ should be solved. 
$\six\alpha_0$ must be solved after we obtain $q_u$ 
because its Poisson equation involves $q_u$ in the source term. 
In Table 2, we also list potentials which are included in the source 
terms of the Poisson equations for other potentials. 

The configuration which we are most interested in and would like to 
obtain is the equilibrium state 
for BNS's of equal mass. Hence, we show the boundary condition at 
$r \rightarrow \infty$ for this problem. 
When we consider equilibrium configurations for 
BNS's where the center of mass for each NS is on the $x$-axis, 
boundary conditions for potentials at $r \rightarrow \infty$ become 
\begin{eqnarray}
U   &=&{1 \over r}\int \rho dV+O(r^{-3}), \hskip 2cm~~~
q_{x}={n^x \over r^2}\int \rho  x^2 dV+O(r^{-4}), \nonumber\\
q_2 &=&{1 \over r}\int \rho R^2 dV+O(r^{-3}),\hskip 2cm
q_{y}={n^y \over r^2}\int \rho  y^2 dV+O(r^{-4}), \nonumber\\
q_e &=&{1 \over r}\int \rho \varepsilon dV+O(r^{-3}),\hskip 2cm~~
q_{z}={n^z \over r^2}\int \rho  z^2 dV+O(r^{-4}), \nonumber\\
q_u&=&{1 \over r}\int \rho U dV+O(r^{-3}),\hskip 2cm~
q_4={1 \over r}\int \rho R^4  dV+O(r^{-3}), 
\end{eqnarray}
\begin{eqnarray}
\five P_{0x}&=&{n^x \over r^2}\int S^{(P)}_x x dV
+{n^y \over r^2}\int S^{(P)}_y y dV +O(r^{-3}), \nonumber\\
\five P_{0y}&=&{n^x \over r^2}\int S^{(P)}_y x dV
+{n^y \over r^2}\int S^{(P)}_y y dV +O(r^{-3}), 
\end{eqnarray}
\begin{eqnarray}
\eta_{0x}&=&{n^x \over r^2}\int \rho x^2\Bigl(\varepsilon+{5 \over 2}U\Bigr)
dV+O(r^{-4}), \nonumber\\
\eta_{0y}&=&{n^y \over r^2}\int \rho y^2\Bigl(\varepsilon+{5 \over 2}U\Bigr)
dV+O(r^{-4}), 
\end{eqnarray}
\begin{eqnarray}
Q^{(I)}_{0x}&=&{n^x \over r^2}\int \rho x^2\Bigl(3(\Gamma-1)\varepsilon
-{1 \over 2}U\Bigr) dV+O(r^{-4}),\hskip 1cm
q_{2x}={n^x \over r^2}\int \rho R^2 x^2 dV+O(r^{-4}), \nonumber\\
Q^{(I)}_{0y}&=&{n^y \over r^2}\int \rho y^2\Bigl(3(\Gamma-1)\varepsilon
-{1 \over 2}U\Bigr) dV+O(r^{-4}),\hskip 1cm
q_{2y}={n^y \over r^2}\int \rho R^2 y^2 dV+O(r^{-4}), \nonumber\\
Q^{(I)}_{0z}&=&{n^y \over r^2}\int \rho z^2\Bigl(3(\Gamma-1)\varepsilon
-{1 \over 2}U\Bigr) dV+O(r^{-4}),\hskip 1cm
q_{2z}={n^z \over r^2}\int \rho R^2 z^2 dV+O(r^{-4}), 
\end{eqnarray}
\begin{eqnarray}
h^{(U)}_{xx}&=&{1 \over r}\int S^{(U)}_{xx}  dV+O(r^{-3}),\hskip 3cm
h^{(U)}_{xy}={3n^x n^y \over r^3}\int S^{(U)}_{xy}xy  dV+O(r^{-5}), \nonumber\\
h^{(U)}_{yy}&=&{1 \over r}\int S^{(U)}_{yy}  dV+O(r^{-3}),\hskip 3cm 
h^{(U)}_{xz}={3n^x n^z \over r^3}\int S^{(U)}_{xz}xz  dV+O(r^{-5}), \\
h^{(U)}_{yz}&=&{3n^y n^z \over r^3}\int S^{(U)}_{yz}yz  dV+O(r^{-5}), \nonumber
\end{eqnarray}
\begin{eqnarray}
q_{xx}&=&{1 \over r}\int \rho x^2  dV+O(r^{-3}),\hskip 3cm
q_{xy}={3n^x n^y \over r^3}\int \rho x^2y^2  dV+O(r^{-5}), \nonumber\\
q_{xz}&=&{3n^x n^z \over r^3}\int \rho x^2 z^2  dV+O(r^{-5}),\hskip 2cm
q_{yz}={3n^y n^z \over r^3}\int \rho y^2z^2  dV+O(r^{-5}),
\end{eqnarray}
\begin{eqnarray}
\six\alpha_0&=&{1 \over r}\int S^{(\alpha_0)}  dV+O(r^{-3}),\hskip 2cm
\six\alpha_2 ={1 \over r}\int S^{(\alpha_2)}  dV+O(r^{-3}),  
\end{eqnarray}
where $dV=d^3x$, and 
\begin{equation}
n^i={x^i \over r}. 
\end{equation}
We note that at $r \rightarrow \infty$, $S_i^{(P)} \rightarrow O(r^{-5})$, 
$S_{ij}^{(U)} \rightarrow O(r^{-6})$, 
$S^{(\alpha_0)} \rightarrow O(r^{-4})$ and 
$S^{(\alpha_2)} \rightarrow O(r^{-4})$, so that 
all the above integrals are well defined.

\section{Conserved quantities}

In this section, we show the conserved quantities in the 2PN 
approximation because they will be useful to investigate the stability 
property of equilibrium solutions obtained in numerical calculations. 

\noindent
(1)Conserved mass\cite{asada}; 
\begin{equation}
M_{\ast} \equiv \int \rho_{\ast}d^3x, 
\end{equation}
where 
\begin{eqnarray}
\rho_{\ast}&=&\rho \alpha u^0 \psi^6  \nonumber\\
&=&\rho \biggl[1+{1 \over c^2}\Bigl({1 \over 2}v^2+3 U\Bigr)
+{1 \over c^4}\Bigl({3 \over 8}v^4+{7 \over 2}v^2 U
+{15 \over 4}U^2+6 \four\psi+\three\beta_i v^i\Bigr)+O(c^{-6})\biggr]. 
\label{stareq}
\end{eqnarray}
Equation $(\ref{stareq})$ may be written as 
\begin{equation}
\rho_{\ast}=\rho \biggl[1+{1 \over c^2}\Bigl({1 \over 2}v^2+3 U\Bigr)
+{1 \over c^4}\Bigl({3 \over 8}v^4+{13 \over 2}v^2 U
+{45 \over 4}U^2+3U\varepsilon+\three\beta_i v^i\Bigr)+O(c^{-6})\biggr].
\end{equation}
\noindent
(2)ADM mass\cite{wald,asada}; 
\begin{equation}
M_{ADM}=-{1 \over 2\pi}\int \Delta \psi d^3 x \equiv 
\int \rho_{ADM}d^3x, 
\end{equation}
where 
\begin{eqnarray}
\rho_{ADM}&=&\rho \biggl[1+{1 \over c^2}
\Bigl(v^2+\varepsilon+{5 \over 2}U\Bigr) 
+{1 \over c^4}\biggl\{v^4+{13 \over 2}v^2 U+\Gamma\varepsilon v^2
+{5 \over 2}U\varepsilon+{5 \over 2}U^2+5\four\psi \nonumber\\
&&~~~~~~~~~~~~~+2\three\beta_i v^i
+{1 \over 32\pi\rho}\three\beta_{i,j}
\Bigl(\three\beta_{i,j}+\three\beta_{j,i}
-{2 \over 3}\delta_{ij}\three\beta_{k,k}\Bigr)
\biggr\} +O(c^{-6}) \biggr], 
\end{eqnarray}
or
\begin{eqnarray}
\rho_{ADM}&=&\rho \biggl[1+{1 \over c^2}
\Bigl(v^2+\varepsilon+{5 \over 2}U\Bigr)
+{1 \over c^4}\Bigl(v^4+9 v^2 U+\Gamma\varepsilon v^2+5 U\varepsilon
+{35 \over 4}U^2 +{3 \over 2}\three\beta_i v^i \Bigr)  \nonumber\\
&& \hskip 9cm +O(c^{-6}) \biggr]. 
\end{eqnarray}
\noindent
(3)Total energy, which is calculated from $M_{ADM}-M_*$ in the third PN 
order\cite{asada}; 
\begin{equation}
E \equiv \int \rho_E d^3x, 
\end{equation}
where 
\begin{eqnarray}
\rho_E &=&\rho\biggl[
\biggl( {1 \over 2}v^2+\varepsilon-{1 \over 2}U \biggr)
+{1 \over c^2}\biggl({5 \over 8}v^4+{5 \over 2}v^2 U
+\Gamma v^2\varepsilon+2 U\varepsilon-{5 \over 2}U^2
+{1 \over 2}\three\beta_i v^i \biggr)  \nonumber\\
&&\hskip 1.5cm 
+{1 \over c^4}\biggl\{ {11 \over 16}v^6+v^4\Bigl(\Gamma \varepsilon
+{47 \over 8} U\Bigr) +v^2 \Bigl( 4\four\psi+6\Gamma \varepsilon U
+{41 \over 8}U^2+{5 \over 2}\three\beta_i v^i-X \Bigr)  \nonumber\\
&&\hskip 2cm
-{5 \over 2}U^3+2\Gamma \three\beta_i v^i \varepsilon+5 \varepsilon\four\psi
+5U \three\beta_i v^i-{15 \over 2}U\four\psi+{5 \over 4}U^2\varepsilon  
\nonumber\\
&&\hskip 2cm +{1 \over 2}h_{ij}v^iv^j
+{1 \over 2}\three\beta_i\three\beta_i  \nonumber\\
&&\hskip 2cm +{U \over 16\pi \rho}\biggl(2h_{ij}U_{,ij}+
\three\beta_{i,j} \Bigl(\three\beta_{i,j}+\three\beta_{j,i}
-{2 \over 3}\delta_{ij}\three\beta_{k,k}\Bigr)
\biggr)\biggr\}+O(c^{-6}) \biggr]. 
\end{eqnarray}

\noindent
(4)Total linear and angular momenta: In the case $K_i^{~i}=0$, 
these are calculated from\cite{wald}
\begin{eqnarray}
P_i&=&{1 \over 8\pi}\lim_{r \to \infty}\oint
K_{ij}n^j dS  \nonumber\\
&=&{1 \over 8\pi}\lim_{r \to \infty}\oint \psi^6
K_{ij} n^j  dS  \nonumber\\
&=&{1 \over 8\pi} \int (\psi^6 K_i^{~j})_{,j} d^3x  \nonumber\\
&=& \int \Bigl( J_i +{1 \over 16\pi} \psi^4 \tilde \gamma_{jk,i}
K^{jk} \Bigr) \psi^6 d^3x, 
\label{jjjeq}
\end{eqnarray}
where $J_i=(\rho c^2+\rho \varepsilon+P) \alpha u^0u_i$. Up to the 
2PN order, the second term in the last line of Eq.$(\ref{jjjeq})$ becomes 
\begin{eqnarray}
&&{1 \over 16\pi} \int  h_{jk,i} \three \beta_{j,k} d^3x, \nonumber\\ 
&=&{1 \over 16\pi} \int \biggl[ \Bigl(h_{jk,i} \three \beta_{j} \Bigr)_{,k}-
h_{jk,ik} \three \beta_{j} \biggr] d^3x, \nonumber\\
&=&{1 \over 16\pi} \lim_{r \to \infty}\oint 
h_{jk,i} \three \beta_{j}  n^k dS=0, 
\end{eqnarray}
where we use $h_{jk} \rightarrow O(r^{-1})$ and  
$\three \beta_{j} \rightarrow O(r^{-2})$ at $r \rightarrow \infty$, and 
the gauge condition $h_{jk,k}=0$. 
Thus, in the 2PN approximation, $P_i$ becomes 
\begin{equation}
P_i \equiv \int p_{i} d^3x ,
\end{equation}
where 
\begin{eqnarray}
p_i&=&\rho \biggl[v^i+{1 \over c^2}\biggl\{
v^i\Bigl(v^2+\Gamma\varepsilon+6U\Bigr)+\three\beta_i \biggr\}
+{1 \over c^4}\biggl\{h_{ij}v^j+\five \beta_i+\three\beta_i\Bigl(v
^2+6U+\Gamma \varepsilon\Bigr) \nonumber\\
&&\hskip 1.5cm+v^i\Bigl(2\three\beta_i v^i+10\four\psi+6\Gamma \varepsilon U+
{67 \over 4}U^2+\Gamma \varepsilon v^2+10Uv^2+v^4-X\Bigr)\biggr\}
+O(c^{-5}) \biggr]. 
\end{eqnarray}
The total angular momentum $J$ becomes
\begin{equation}
J= \int p_{\varphi} d^3x , 
\end{equation}
where $p_{\varphi}=-yp_x+xp_y$. 

\section{Summary}

It is generally expected that there exists no Killing vector in the 
spacetime of coalescing BNS's because such a spacetime is filled with 
gravitational radiation which propagates to null infinity. However, we may 
consider coalescing BNS's as the almost stationary object 
from physical point of view as described in section 1. 
Motivated by this idea, in this paper, 
we have developed a formalism to obtain equilibrium 
configurations of uniformly rotating fluid up to the 2PN order 
using the PN approximation.  The concept of being ``almost'' stationary  
becomes clear in the framework of the PN approximation and,   
in particular, the stationary rotating objects 
can exist exactly at the 2PN order, since the energy loss due to 
the gravitational radiation does occur from the 2.5PN order. 
There appear, at the 2PN order, tensor potentials $h_{ij}$   
which were completely ignored in Wilson's approach\cite{wilson}. 
It should be noted that these tensor potentials play an important role 
at the 2PN order: This is because 
they appear in the equations to 
determine equilibrium configurations as shown in previous sections and 
they also 
contribute to the total energy and angular momentum of systems. 
This means that if we performed the stability analysis ignoring the tensor 
potentials, we might reach an incorrect conclusion. 

In our formalism, we extract terms depending on the angular velocity $\Omega$ 
from the integrated 
Euler equation and Poisson equations for potentials, and rewrite 
the integrated Euler equation as an explicit equation in $\Omega$. 
This reduction will improve the convergence in numerical 
iteration procedure. 
As a result, the number of Poisson equations we need to solve in each step 
of iteration reaches 29. 
However, source terms of the Poisson equations 
decrease rapidly enough, at worst $O(r^{-4})$, 
in the region far from the source, so that we can solve accurately 
these equations as the boundary value problem like in the case of 
the first PN calculations\cite{shibata}.  
Thus, the present formalism will be useful to obtain equilibrium 
configurations for synchronized BNS's or the Jacobi ellipsoid. 
These configurations will be obtained in future work. 

\vskip 5mm

{\bf Acknowledgments}

For helpful discussions, 
we would like to thank T. Nakamura, M. Sasaki and T. Tanaka. 
H. A. would like to thank Professor S. Ikeuchi, Professor M. Sasaki 
and Professor Futamase for their encouragement. 
This work was in part supported by the Japanese Grant-in-Aid on 
Scientific Research of the Ministry of Education, Science, and Culture, 
No. 07740355.

\vspace{5mm}


\vspace{3cm}

\begin{center}
{\bf Figure Captions}
\end{center}

\begin{itemize}
\item[{\rm Fig}.~1] Error of the pressure in the post-Newtonian 
approximation for the GR compact star of uniform density as a function 
of the normalized areal radius($r/R$). Solid and dotted lines show the case 
$R=5Gm/c^2$ and $8Gm/c^2$, where $R$ and $m$ are the circumference raduis 
and the mass of star, respectively. 
\end{itemize}

\centerline{}

\newpage

\begin{center}
{\bf Table 1 (a)}
\end{center}
\noindent
Various levels of approximation in terms of PN expansions$(v^2/c^2)$ and 
mass ratio($\eta=\mu/M$; $\mu=$ reduced mass, $M=$ total mass). 
We mark $P^2N$ if all terms in that level are taken into account in the 
2PN approximation, while $W$ is marked if Wison's approach takes 
into account all terms in that level. 
The mark $-$ means that the relevant term does not exist and 
the levels taken into account by neither approaches are blank. 
We neglect secular effects due to gravitational radiation reaction 
in Tables 1(a) and (b). 
It should be noted that, at $O(\eta^0)$, Wilson's approach produces 
exact GR solutions, but it is not justified at the 2PN order 
even at $O(\eta^1)$. 

$$\vbox{\offinterlineskip
\tabskip=0pt
\def\tablerule{\noalign{\hrule}}
\def\tablespace{\omit&height2pt&\omit&&\omit&&\omit&&\omit&&\omit&\cr}
\halign{
\strut#&\vrule#&
\quad\hfil#\hfil\quad &\vrule#& \quad\hfil#\hfil\quad &\vrule#&
\quad\hfil#\hfil\quad &\vrule#& \quad\hfil#\hfil\quad &\vrule#&
\quad\hfil#\hfil\quad &\vrule#& \quad\hfil#\hfil\quad &\vrule#&
\quad\hfil#\hfil\quad &\vrule#& \quad\hfil#\hfil\quad &\vrule# \cr
\tablerule\tablespace
&& PN$~\setminus~$$\eta$ && $\eta^0$ && $\eta^1$ && $\eta^2$ && $O(\eta^3)$ 
& \cr\tablespace\tablerule\tablespace
&& N && $P^2N$, $W$ && $-$ && $-$ && $-$
& \cr\tablespace\tablerule\tablespace
&& 1PN && $P^2N$, $W$ && $P^2N$, $W$ && $-$ && $-$
& \cr\tablespace\tablerule\tablespace
&& 2PN && $P^2N$, $W$ && $P^2N$ && $~~~P^2N~~~$ && $~~~~~-~~~~~$
& \cr\tablespace\tablerule\tablespace
&& $\ge$3PN && $W$ &&   &&   &&
& \cr\tablespace\tablerule
}}$$

\begin{center}
{\bf Table 1 (b)}
\end{center}

\noindent
Various levels of approximation in terms of PN expansions$(Gm/c^2R)$ and 
ellipticity of a NS$(e)$. 
The meanings of $P^2N$ and $W$ are the same as those in Table 1(a).
Wilson's approach produces exact GR solutions in the case of the completely 
spherical star.  

$$\vbox{\offinterlineskip
\tabskip=0pt
\def\tablerule{\noalign{\hrule}}
\def\tablespace{\omit&height2pt&\omit&&\omit&&\omit&\cr}
\halign{
\strut#&\vrule#&
\quad\hfil#\hfil\quad &\vrule#& \quad\hfil#\hfil\quad &\vrule#&
\quad\hfil#\hfil\quad &\vrule#& \quad\hfil#\hfil\quad &\vrule#&
\quad\hfil#\hfil\quad &\vrule#& \quad\hfil#\hfil\quad &\vrule#&
\quad\hfil#\hfil\quad &\vrule#& \quad\hfil#\hfil\quad &\vrule# \cr
\tablerule\tablespace
&& PN$~\setminus~$ $e$ && $e=0$ && $e\neq 0$
& \cr\tablespace\tablerule\tablespace
&& N && $P^2N$, $W$ && $P^2N$, $W$
& \cr\tablespace\tablerule\tablespace
&& 1PN && $P^2N$, $W$ && $P^2N$, $W$
& \cr\tablespace\tablerule\tablespace
&& 2PN && $P^2N$, $W$ && $P^2N$
& \cr\tablespace\tablerule\tablespace
&& $\ge$3PN && $W$ && 
& \cr\tablespace\tablerule
}}$$

\thispagestyle{empty}

\newpage

\thispagestyle{empty}

\begin{center}
{\bf Table 2}
\end{center}

\noindent
List of potentials to be solved(column 1), 
Poisson equations for them(column 2), 
and other potential variables which appear in the source term of the 
Poisson equation(column 3). Note that 
$i$ and $j$ run $x,y,z$. Also, 
note that we do not have to solve $\eta_{0z}$, $\five P_{0z}$, 
$q_{yy}$, $q_{zz}$ and $h_{zz}^{(U)}$. 

$$\vbox{\offinterlineskip
\tabskip=0pt
\def\tablerule{\noalign{\hrule}}
\def\tablespace{\omit&height2pt&\omit&&\omit&&\omit&&
\omit&&\omit&&\omit&&\omit&\cr}
\halign{
\strut#&\vrule#&
\quad\hfil#\hfil\quad &\vrule#& \quad\hfil#\hfil\quad &\vrule#&
\quad\hfil#\hfil\quad &\vrule#& \hfil#\hfil~ &\vrule#&
\quad\hfil#\hfil\quad &\vrule#& \quad\hfil#\hfil\quad &\vrule#&
\quad\hfil#\hfil\quad &\vrule#& \quad\hfil#\hfil\quad &\vrule#&
\quad\hfil#\hfil\quad &\vrule#& \quad\hfil#\hfil\quad &\vrule# \cr
\tablerule\tablespace
&& Pot. && Eq. && Needed pots. && && Pot. && Eq. && Needed pots. 
& \cr\tablespace\tablerule\tablespace
&& $U$ && (2.11) && None && && $q_{ij}$ && (4.6) && None  
& \cr\tablespace\tablerule\tablespace
&& $q_i$ && (3.14) && None && && $Q_{0i}^{(I)}$ && (4.15) && $U$
&\cr\tablespace\tablerule\tablespace
&& $q_2$ && (4.1)&& None && && $\eta_{0i}$ && (4.16) && $U$
&\cr\tablespace\tablerule\tablespace
&& $q_{2i}$ && (4.2) && None && && $\five P_{0i}$ && (4.17) && $U,~q_i$
&\cr\tablespace\tablerule\tablespace
&& $q_4$ && (4.3) && None && && $\six\alpha_0$ && (4.21) 
&& $U,~q_e,~q_u,~h_{ij}^{(U)},~Q_{0i}^{(I)}$
&\cr\tablespace\tablerule\tablespace
&& $q_u$ && (4.4) && $U$  && && $\six\alpha_2$ && (4.22) 
&& $U,~q_2,~q_i,~q_{2i},~q_{ij}$
&\cr\tablespace\tablerule\tablespace
&& $q_e$ && (4.5) && None && && $h_{ij}^{(U)}$ && (3.1) && $U$
&\cr\tablespace\tablerule
}}$$

\newpage

\thispagestyle{empty}

\begin{center}
{\bf Table 3}
\end{center}

\noindent
Variables to be solved in order to obtain the original metric variables. 

$$\vbox{\offinterlineskip
\tabskip=0pt
\def\tablerule{\noalign{\hrule}}
\def\tablespace{\omit&height2pt&\omit&&\omit&&\omit&\cr}
\halign{
\strut#&\vrule#&
\quad\hfil#\hfil\quad &\vrule#& \quad\hfil#\hfil\quad &\vrule#&
\quad\hfil#\hfil\quad &\vrule#& \quad\hfil#\hfil\quad &\vrule#&
\quad\hfil#\hfil\quad &\vrule#& \quad\hfil#\hfil\quad &\vrule#&
\quad\hfil#\hfil\quad &\vrule#& \quad\hfil#\hfil\quad &\vrule# \cr
\tablerule\tablespace
&& Metric && Variables to be solved && see Eq.
& \cr\tablespace\tablerule\tablespace
&& $U$ && $U$ && (2.11)
& \cr\tablespace\tablerule\tablespace
&& $\three\beta_i$ && $q_i$,~$U$ && (3.17)
& \cr\tablespace\tablerule\tablespace
&& $X$ && $q_2$,~ $q_u$,~ $q_e$ && (4.7)
& \cr\tablespace\tablerule\tablespace
&& $\four\psi$ && $q_2$,~ $q_u$,~ $q_e$ && (4.8)
& \cr\tablespace\tablerule\tablespace
&& $\five \beta_i^{(A)}$ && $\five P_{0i}$,~$\eta_{0i}$,~$q_u$,~$q_e$ &&(4.18)
& \cr\tablespace\tablerule\tablespace
&& $\five \beta_i^{(B)}$ && $q_{2i}$,~$q_2$ && (4.18) 
& \cr\tablespace\tablerule\tablespace
&& $\six\alpha$ &&  $\six\alpha_0$,~ $\six\alpha_2$,~ $q_4$ &&(4.20)
& \cr\tablespace\tablerule\tablespace
&& $h_{ij}^{(U)}$ && $h_{ij}^{(U)}$ && (3.1)
& \cr\tablespace\tablerule\tablespace
&& $h_{ij}^{(S)}$ && $q_{ij},q_2$ &&(4.14)
& \cr\tablespace\tablerule\tablespace
&& $h_{ij}^{(A)}$ && $Q_{0i}^{(I)}$,~ $q_u$,~ $q_e$ && (4.19)
& \cr\tablespace\tablerule\tablespace
&& $h_{ij}^{(B)}$ && $q_{ij}$,~ $q_2$,~ $q_{2i}$,~ $q_i$ && (4.19)
& \cr\tablespace\tablerule
}}$$


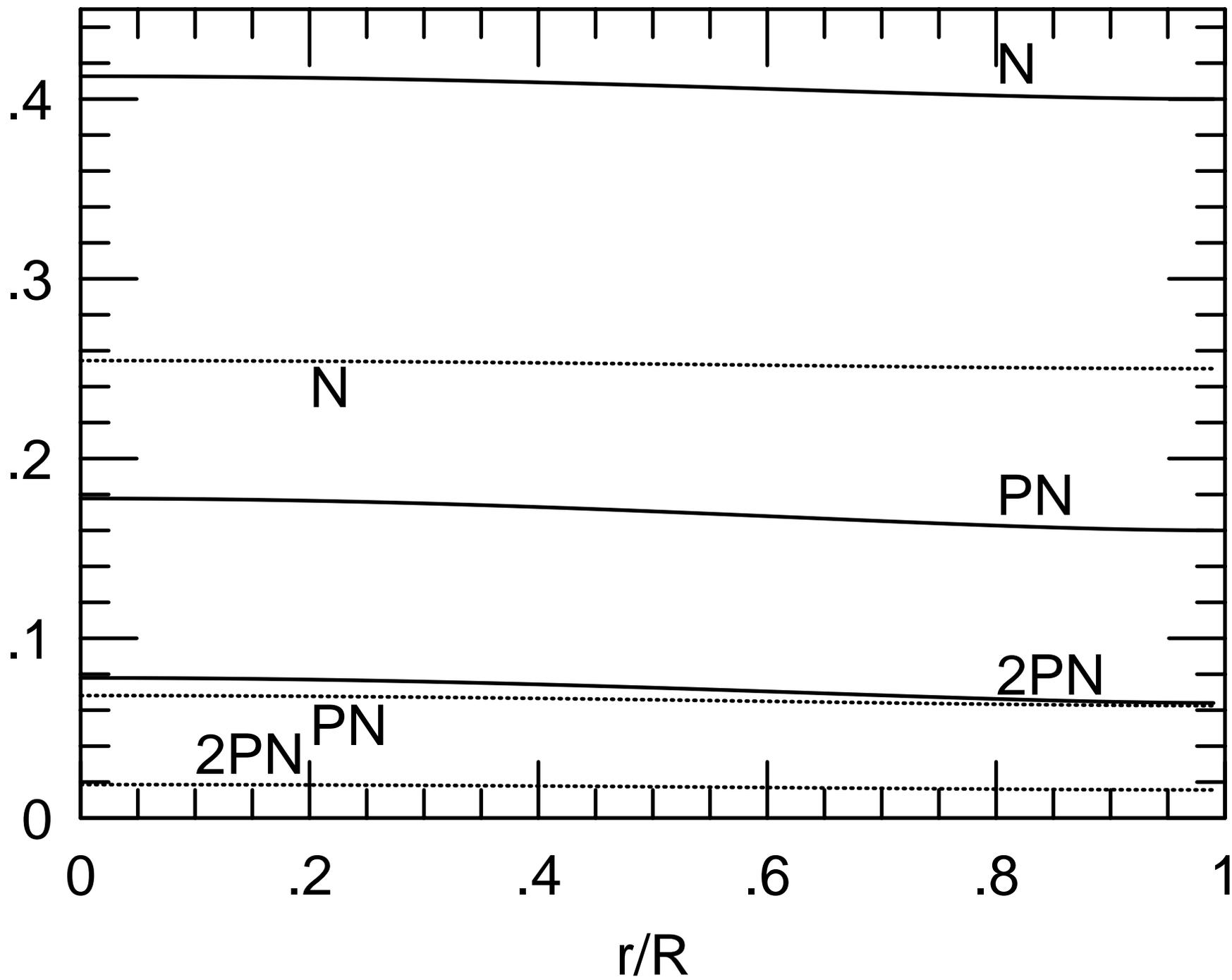
\end{document}